\documentclass[aps,prl,twocolumn,superscriptaddress,nofootinbib]{revtex4-1}
\usepackage{graphicx}
\usepackage{mathtools}
\usepackage{braket}
\usepackage{xcolor}

\newcommand{\Var}[1]{\mathrm{Var}(#1)}

\usepackage{amsmath}
\usepackage{amsfonts,dsfont}
\newcommand{\Varr}[0]{\mathrm{Var}}
\newcommand*{\defeq}{=}

\begin{document}
\title{Quantum Absorbance Estimation and the Beer-Lambert Law}
\author{Euan J. Allen}\email{euan.allen@bristol.ac.uk}
\affiliation{Quantum Engineering Technologies Labs, H. H. Wills Physics Laboratory and Department of Electrical \& Electronic Engineering, University of Bristol, BS8 1FD, United Kingdom}
\affiliation{Quantum Engineering Centre for Doctoral Training, Nanoscience and Quantum Information Centre, University of Bristol, BS8 1FD, United Kingdom}
\author{Javier Sabines-Chesterking}
\affiliation{Quantum Engineering Technologies Labs, H. H. Wills Physics Laboratory and Department of Electrical \& Electronic Engineering, University of Bristol, BS8 1FD, United Kingdom}
\author{Alex McMillan}
\affiliation{Quantum Engineering Technologies Labs, H. H. Wills Physics Laboratory and Department of Electrical \& Electronic Engineering, University of Bristol, BS8 1FD, United Kingdom}
\author{Siddarth K. Joshi}
\affiliation{Quantum Engineering Technologies Labs, H. H. Wills Physics Laboratory and Department of Electrical \& Electronic Engineering, University of Bristol, BS8 1FD, United Kingdom}
\author{Peter S. Turner}
\affiliation{Quantum Engineering Technologies Labs, H. H. Wills Physics Laboratory and Department of Electrical \& Electronic Engineering, University of Bristol, BS8 1FD, United Kingdom}
\author{Jonathan C. F. Matthews}
\affiliation{Quantum Engineering Technologies Labs, H. H. Wills Physics Laboratory and Department of Electrical \& Electronic Engineering, University of Bristol, BS8 1FD, United Kingdom}

\begin{abstract}
The utility of transmission measurement has made it a target for quantum enhanced measurement strategies. Here we find if the length of an absorbing object is a controllable variable, then via the Beer-Lambert law, classical strategies can be optimised to reach within 83\% of the absolute quantum limit. Our analysis includes experimental losses, detector noise, and input states with arbitrary photon statistics. We derive optimal operating conditions for both classical and quantum sources, and observe experimental agreement with theory using Fock and thermal states. %This work supports the need to focus quantum measurement resources on scenarios where a clear and substantial quantum advantage is achievable.

%Transmission measurements are ubiquitous in metrology. Most often the measured transmission depends on well known constants, such as apparatus loss, %(e.g., due to the measurement apparatus, sample size or impurities) 
%and the sample being measured. In such cases, both classical and quantum optimisations are possible. In a case study of Beer-Lambert law we show a 20\% quantum advantage despite classical optimisations.
%Our case study of the Beer-Lambert law proves experimentally and theoretically that a quantum measurement has upto a 20\% advantage even when compared with a classically optimised measurement. 
%Our analysis covers experimental noise \& losses and input states with an arbitrary Fano factor. We derive optimal operating conditions for both classical and quantum sources and observe agreement between experiment and theory using Fock and thermal states.
%The Beer-Lambert law describes the attenuation of light through a length $L$ of absorbing material with absorbance $a$ and has wide application across science and engineering. Here, we demonstrate the achievable precision of estimating $a$ with quantum light, both theoretically and experimentally. We find for cases where the experiment is performed with an optimal choice of $L$, the advantage provided by quantum states is bounded to 20\%. We include in our analysis experimental losses, states of arbitrary Fano factor, and detector dark counts. We derive optimal operating conditions for both classical and quantum sources and observe agreement between experiment and theory using Fock and thermal states.
\end{abstract}

\maketitle

% general structure
% INTRO
% Laser noise is an issue
% Here are all the applications that care about laser noise
% Here are ways you can reduce laser noise
% Here is why they might not be great
% AMZI
% new theory
% Done in free space, fibre good/works well (supp. mat.) but has significant draw backs on reaching SNL due to raman/loss/disp/spm. 
% hollow core fibre
% works well/SNL
% Dbl up

%\section{Introduction}
In sensing scenarios, a measured parameter is often a function of known constants of the experiment and the particular variable of interest. An example of this is the Beer-Lambert law, where the intensity of radiation transmitted through a sample of length $L$ is given by: $I =~I_0 \exp{(-a L)}$, where $I_0$ and $I$ are the intesities before and after the sample, and $a$ is the sample absorbance parameter. The law is ubiquitous across a wide range of investigative techniques including atomic vapour thermometry~\cite{truong2011quantitative}, femtosecond pump-probe spectroscopy~\cite{woutersen1997femtosecond}, high-throughput screening~\cite{inglese2007high}, on- and in-line food processing~\cite{scotter1990use}, medical diagnostics~\cite{behera2012uv}, and spectrophotometry~\cite{trumbo2013applied}. The Beer-Lambert law also applies to non-optical techniques such as neutron transmission and electron tomography~\cite{vontobel2006neutron,yan2015fast}.

Optical quantum metrology investigates how quantum strategies, such as probing with quantum states of light, provides increased precision over classical techniques in estimating parameters including
%. There have been a number of theoretical investigations
%%into viable sensing schemes 
%covering a range of 
%sensing 
%parameters of interest such as 
optical phase~\cite{caves1981quantum,ono2013entanglement}, transmission~\cite{javier2017sub,moreau2017demonstrating}, polarisation~\cite{zhang2016universal}, and displacements~\cite{braginsky1967classical}. A variety of practical implementations have demonstrated these schemes and have been shown to outperform classical strategies operating at the same average input intensity~\cite{aasi2013enhanced}. 
%\jm{\textit{$<$Combine expert' \& theory?$>$}}

%T
Motivated by the utility 
%and simplicity of using 
of measuring optical absorption to image and identify objects, there has 
%been considerable
been a series of studies exploring
%work looking at 
the benefits of using quantum 
%states of 
light to estimate
%ion of a
%a 
the total transmission 
%parameter 
$\eta$, where the average intensity passing through the loss channel is ${I} = \eta {I}_0$~\cite{fujiwara2004estimation,adesso2009optimal,brida2010experimental,moreau2017demonstrating}. Here, we investigate how the advantage of applying optimal quantum states 
%of light 
changes, if the parameter sought
%you are trying to estimate 
is the \emph{absorbance} coefficient $a$ (loss per unit length) used in the Beer-Lambert law~\cite{whittaker2017absorption}. %We note that $a$ and $\eta$ are different experimental parameters and so are not directly comparable. %Here we investigate if estimating $a$ is improved in a similar fashion by applying quantum probes.

%In this paper we theoretically investigate quantum absorbance estimation. We first begin with an ideal lossless implementation and discuss the advantage offered over classical strategies. We show that optimising of the length f absorbing material can drastically reduce the advantage provided by quantum sensing strategies. We move on to discuss non-ideal implementations where there is experimental loss present or cases where the classical or quantum input are not operating with the optimal noise properties. We show that although these alterations change the procedure slightly, optimisation over the length to be performed. We demonstrate how this optimising is analogous to previous work investigating multipass strategies for phase and loss estimation. We end with an experimental demonstration of the theoretical work of this paper and demonstrate how dark counts affect the acquired result. 

\begin{figure}[hbtp]
\centering
\includegraphics[width=0.8\linewidth]{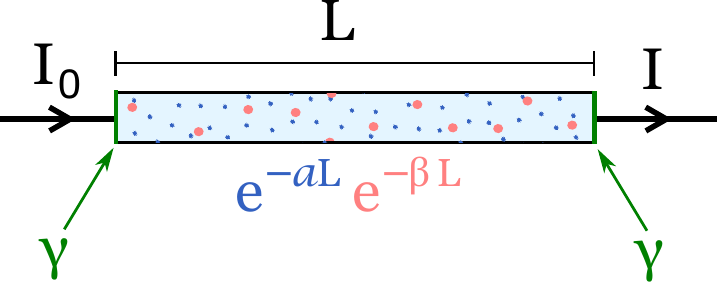}
\caption{A typical experiment to measure the absorbance of a medium in the presence of other loss mechanisms such as surface losses $\gamma$ and co-propagation loss $\beta$. 
%The mean intensity at the output $I_{out}$ from a sample with incident intensity $I_{in}$ and transmission $\eta = e^{-a L}$, where $a$ is the absorbance and $L$ is the length of the medium, is given by $I_{out} = \eta I_{in}$. 
\label{fig:theoryExptsetup}}
\end{figure}

%\section{Theory}
%\subsection{Quantum Absorbance Estimation}
The 
%experimental schematic we will be considering for the following analysis 
scenario we consider
%is shown in 
(Fig.~\ref{fig:theoryExptsetup}) comprises the targeted
%. The 
absorbance $a \in (0, \infty)$, a variable length $L \in (0, \infty)$, length independent loss $\gamma$ which, for example, could arise from loss at each facet of the sample chamber, and a length dependent loss per unit length $\beta$, which we refer to as co-propagating loss, and could arise from a distinct absorbing material in the chamber with the sample. 
%have domains that 
%provide a corresponding range of absorption values $\eta = \exp(-a L) \in (0, 1)$ consistent with discussions of quantum absorption estimation~\cite{whittaker2017absorption}. 
%We \jm{also} include in \jm{our}
%%this
%analysis 
%%not only the loss provided by the sample of interest, but also 
%%experimental loss caused by practical challenges. In particular, we look at\jm{:}
%%two types of loss: 
%%1) 
%length independent loss \jm{$\gamma$} which, for example, could arise from \jm{loss at each} facet 
%%loss 
%\jm{of}
%%into 
%the sample chamber, and 
%%2) 
%length dependent 
%%or 
%co-propagating loss \jm{per unit length $\beta$}, which could arise from another type of \jm{absorbing} material in the chamber with the sample. \jm{Input intensity \jm{$I_0$} then becomes to output \jm{$I$} by}
%or from the chamber itself having intrinsic loss\footnote{For example, this could occur if the light is propagating in a slot waveguide structure that has propagation loss even if the material of interest is not present.}.
%Including both 
%%of these 
%loss mechanisms \jm{relates}
%%results in the following relationship between 
%input \jm{$I_0$} and output \jm{$I$} intensities \jm{by}
Input intensity $I_0$ is then related to output $I$ by
\begin{equation}
I = I_0 \exp(-a L) \eta_l,
\end{equation}
where we group the instrumental `non-sample' loss mechanisms as $\eta_l=\gamma^2 \exp(-\beta L)$ and facet loss is assumed to be the same for entrance and exit to simplify expressions.
%where 
%%$I, I_0, a, L$ are as before, 
%$\beta$ is 
%%the 
%co-propagating loss per unit length and $\gamma$ is 
%%the 
%facet transmission, applied twice for \jm{each facet}
%%the front and back facets 
%(which we assume for brevity to \jm{be} 
%%have 
%the same). 
%loss). 
%In some cases w
%\jm{W}e group 
%\jm{these external}
%%all experimental 
%loss mechanisms \jm{as}
%%under 
%%a single parameter 
%$\eta_l=\gamma e^2 \exp(-b L)$ to simplify %the 
%expressions. 
%In our model, we assume 
%that 
Variables $L$, $\beta$, and $\gamma$ are assumed to be known a priori to infinite precision 
%prior to the experiment 
and so contribute no uncertainty to 
estimating
%the estimate of 
$a$. We note that both length dependent loss variables $a$ and $\beta$ are multiplied by the same length $L$ as any loss occurring outside of $L$ can be encapsulated in the parameter $\gamma$.

 %The metric \jm{we use} 
%used in the following analysis 
%t
To compare different 
experimental strategies, we use
%is 
the Fisher information per average incident photon into the sample, $\mathcal{F(x)}=F(x)/N_0$, where $F(x)$ is the total Fisher information for the probe state on the parameter $x$ and $N_0$ is the mean input photon number. 
%Completing the following analysis for 
Using Fisher information per absorbed photon, $F/(N_0(1-\eta))$, does not 
%change 
alter the 
qualitative
%occurrence of the 
results observed
%described in this paper 
but can alter
%cause slight differences in 
the optimal numerical values of 
%certain 
parameters (which can still be analytically defined). 
%For the rest of this paper, the Fisher information on a parameter $x$ per incident photon will be marked in equations as $\mathcal{F}(x)$.

\begin{figure*}[hbtp]
\centering
\includegraphics[width=\linewidth]{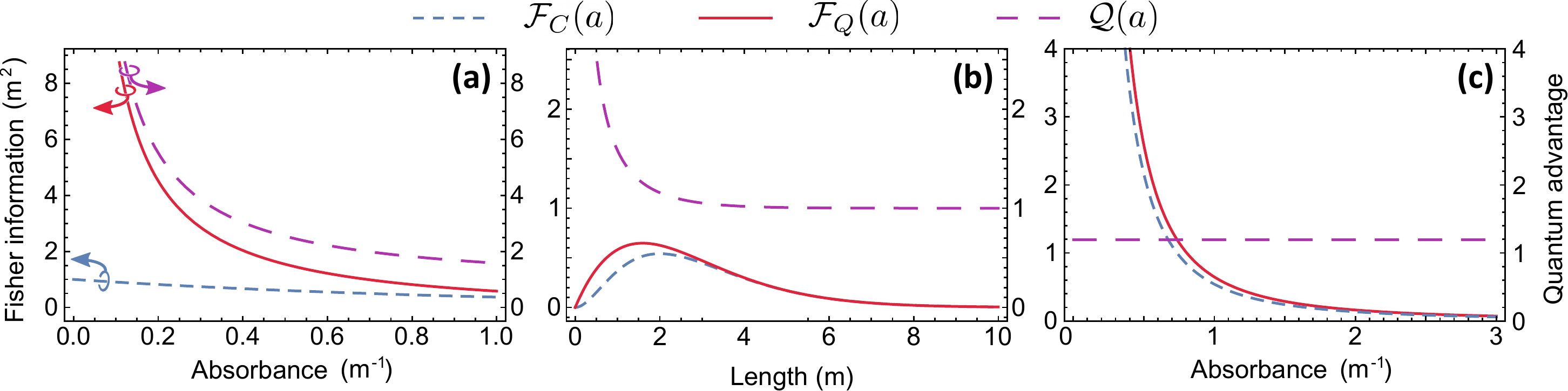}
\caption{
(a) 
%Fisher information on the absorbance $a$ per incident photon 
$\mathcal{F}(a)$ given a fixed 
%sample length of 
$L=1$ 
%(
for both classical and quantum strategies.
%). 
Also shown is the quantum advantage $\mathcal{Q}(a)$ (ratio of quantum to classical Fisher information). The trend follows that previously found from absorption estimation, where weakly absorbing samples are most improved by using quantum states of light. Colour arrows dictate relevant axis for each line.
(b) $\mathcal{F}(a)$ and quantum advantage $\mathcal{Q}(a)$ given a fixed sample absorbance of $a=1$. We see that the Fisher information in both classical and quantum cases peaks at particular length values. 
(c) $\mathcal{F}(a)$ and quantum advantage $\mathcal{Q}(a)$ when both schemes are allowed to operate at the optimal length values given by Eq.~\ref{eq:optimallengthC} and~\ref{eq:optimallengthQ}. We see that maximising the Fisher information in the classical and quantum case means that the quantum advantage is fixed to a value of $ 1.2$ for all values of absorbance.\label{fig:FImultiplots}}
\end{figure*}

For estimation of the total absorption $\eta$,
% it has previously been found~\cite{fujiwara2004estimation,adesso2009optimal} that 
the best known quantum strategy is to input Fock states, $\ket{N_0}$,  of known photon number $N_0$ into the sample, and measure the output intensity~\cite{fujiwara2004estimation,adesso2009optimal}. The corresponding classical strategy uses
%involves an equivalent measurement with 
a coherent state $\ket{\alpha}$, with mean photon number $N_0 = | \alpha |^2$. While absorbance and absorption estimation differ in parametrisation, in both cases the 
%fundamental 
underlying
evolution of any input quantum state is the same physical process of a
%is still equivalent to a 
loss channel. 
%As the physical process is the same in both cases, 
Therefore the Fisher information of absorption and absorbance are proportional to one another (Eq.~\ref{eq:BLlaw_reparamFI})~\cite{lehmann2006theory,paris2009quantum}, and so share the same optimal quantum and classical strategies.
%. Because of this we expect no change in the optimality of quantum or classical states of light in both cases as maximising the value in one case finds the maximum for both.
%These strategies continue to be optimal for absorbance estimation and hence 
%w
We 
%will 
therefore can consider the difference between these two input states as defining the difference between classical ($\ket{\alpha}$ input) and quantum ($\ket{N_0}$ input) strategies in absorbance estimation. 

Because
%With 
$\eta=\exp(-a L)$
%then $\eta$
is a continuous differentiable function of $a$,
%and therefore 
the relationship relating the Fisher information on each parameter applies~\cite{lehmann2006theory}:
\begin{eqnarray}
\mathcal{F}(a) &= \left( \frac{\partial \eta}{\partial a} \right)^2 \mathcal{F}(\eta) \nonumber \\ 
&= L^2 e^{-2aL} \mathcal{F}(\eta). \label{eq:BLlaw_reparamFI}
\end{eqnarray}
%Previously it has been found\jm{~\cite{Ref??}} that f
For absorption estimation with 
%experimental loss present, 
$\eta_l$ present, for classical and quantum strategies, we have~\cite{supplmaterial}
%the Fisher information per incident photon on $\eta$ for classical and quantum strategies \jm{are~\cite{Ref??}}
%goes as
\begin{eqnarray}
&\mathcal{F}_C(\eta) =  \frac{\eta_l}{\eta}  = \eta_l e^{aL}, \label{eq:FishClass}\\
&\mathcal{F}_Q(\eta) = \frac{\eta_l}{\eta (1- \eta \eta_l)} = \frac{\eta_l e^{aL}}{1-\eta_l e^{-aL}} \label{eq:FishQuant}.
\end{eqnarray}
%respectively 
 When combined with Eq.~\ref{eq:BLlaw_reparamFI}, 
%these equations 
Eqs.~\ref{eq:FishClass}, \&~\ref{eq:FishQuant}
provide 
%the following values for the 
Fisher information per incident photon for estimating $a$
\begin{align}
\mathcal{F}_C(a)  = L^2  \gamma^2 e^{-(a+\beta) L} , \label{eq:qsensing_BLfisherinfoC} \\
\mathcal{F}_Q(a) = \frac{L^2  \gamma^2 }{e^{(a+\beta) L} - \gamma^2}. \label{eq:qsensing_BLfisherinfoQ}
\end{align}
These are plotted 
%functions are show 
in Fig.~\ref{fig:FImultiplots}(a) for a fixed $L=1$ and no experimental loss $\eta_l = 1$, along with the quantum advantage $\mathcal{Q}(a) = \mathcal{F}_Q(a) / \mathcal{F}_C(a)$.
We see that fixing $L$
%Fixing the length in Fig.~\ref{fig:FImultiplots}(a) 
provides a scaling for 
%the Fisher information 
$\mathcal{Q}(a)$ that follows the trend
%. As we can see, the trend of the quantum and classical Fisher information (and therefore the quantum advantage) follow similar trends for the case 
of absorption estimation, namely that $\mathcal{Q}(a) \rightarrow \infty$ for $a\rightarrow0$ ($\mathcal{Q}(\eta)\rightarrow\infty$ for $\eta\rightarrow1$)~\cite{javier2017sub}. 
%It is the case that samples that offer the lowest absorption or absorbance value are most improved by applying quantum states of light at the input.

%Similarly, w
We 
%can 
see in Fig.~\ref{fig:FImultiplots}(b) how $\mathcal{F}(a)$
%the Fisher information 
changes for a fixed sample absorbance ($a = 1$), no experimental loss $\eta_l =1$ and a varying 
%length
$L$. This behaviour differs from Fig.~\ref{fig:FImultiplots}(a).
%. This is shown in Fig.~\ref{fig:FImultiplots}(b) and demonstrates different behaviour than the case of Fig.~\ref{fig:FImultiplots}(a). 
For a varying length we see that both $\mathcal{F}_C(a)$ and $\mathcal{F}_Q(a)$ have maximal values before tending towards zero for both large and small $L$. This demonstrates that for a particular 
%absorbance value
$a$ (e.g. a particular gas fixed in concentration), optimising the length of the medium that the light passes through 
%will 
provides maximum information on 
%the parameter 
$a$. For the classical Fisher information provided in Eq.~\ref{eq:qsensing_BLfisherinfoC}, the optimum can be found by solving ${\partial \mathcal{F}_C(a)}/{\partial L} =0$:
\begin{align}
\frac{\partial \mathcal{F}_C(a)}{\partial L} &= 2L \gamma^2 e^{-(a+\beta)L} - (a+\beta)L^2 \gamma^2 e^{-(a+\beta)L} \nonumber \\
&= (2-(a+\beta)L)L \gamma^2 e^{-(a+\beta)L} = 0,
\end{align}
which for non-zero $L$ and $a$, is only satisfied for $(a+\beta)L = 2$. Therefore for a coherent state 
%the optimal value for $L$, providing the maximum Fisher information, is given by
\begin{equation}
L_{\text{opt}}^C = {2}/{(a+\beta)},\label{eq:optimallengthC}
\end{equation}
maximises $\mathcal{F}_C(a)$. For
%which for 
$a=1$, $\beta=0$, and $\gamma=1$, this is in agreement with Fig.~\ref{fig:FImultiplots}(b). An identical process for 
$\mathcal{F}_Q(a)$
%the quantum Fisher information 
%(using Eq.~\ref{eq:qsensing_BLfisherinfoQ}) finds that
yields the optimal
\begin{equation}
L_{\text{opt}}^Q = \frac{\mathcal{W} \left[ - 2 \gamma^2 / e^2 \right]+2}{a+\beta},\label{eq:optimallengthQ}
\end{equation} 
where $\mathcal{W}[x]$ is the principal value of the Lambert W-function~\cite{corless1996on}. As the values of $L_{\text{opt}}$ are inversely proportional to 
%the absorbance 
$a$, the optimal lengths correspond to constant total absorption values. For $\beta=0$ and $\gamma=1$, these are $\eta_{\text{optimal}}^Q = 0.20$ and $\eta_{\text{optimal}}^C = 0.14$. This shows that for any fixed $a$, $L$
%a length of absorbing material 
should be chosen to
%that 
provide
%s around 
approximately $\geq$80\% total absorption through the sample. 

We now compare $\mathcal{F}_C(a)$ and $\mathcal{F}_Q(a)$
%the Fisher information of classical and quantum schemes 
when both schemes are allowed 
%sit at 
their optimal $L_{\text{opt}}$ value.
% for each respective input. 
We compute the quantum advantage as a function of absorbance where at each value of $a$ the length of the material is set to $L_{\text{opt}}(a)$, and arrive at Fig.~\ref{fig:FImultiplots}(c). Here we see that in the case where there are no constraints on $L$, the advantage gained by using quantum states of light is limited to a fixed factor of $\mathcal{Q}(a) = 1.2$ for any 
%value of 
$a$. Alternatively, this shows that classical light can reach to within $83\%$ of the fundamental quantum bound. This shows 
%that the effect of having 
free choice of the length parameter $L$ can severely limit the benefit of implementing a quantum strategy for estimating the absorbance. This effect can also be seen analytically by inputting the respective values for $L_{\text{opt}}$ into the expression for
%equation for the 
quantum advantage. 

An important feature of 
%the optimal absorbance length 
$L_{\text{opt}}$
%values 
is that
for
%, in 
both strategies
%cases, 
it is a function of the absorbance.
%it is a function of the absorbance $a$. 
Since $a$ is the parameter being estimated, there are cases where it will not be known in advance. 
%This is the parameter that is supposed to be estimated by implementation of the optimal length, and so in some cases will not be known in advance. 
There do, however, exist 
%number of 
practical
scenarios
% cases 
 when one may still be able to implement $L_{\text{opt}}$.
% optimal length. 
% The first 
One is when the experiment is intended
% designed 
 to measure a deviation from an initially known $a$ by $\delta a$, such as
%in absorbance value, 
% rather than an absolute value 
% (e.g. 
 a change in the concentration of a gas.
% \jm{---here}
% . Here 
% the \jm{initial} value of $a$ is known beforehand and the experimentalist wants to know when this deviates by some amount $\delta a$ which signifies a change in the experiment. 
 Another scenario is
% case might be 
 when $a$ is known approximately and the quantum strategy is employed to
% experiment is to 
 achieve a more precise measurement of its value. Finally, in cases when $a$ is completely unknown beforehand but $L$ can be easily varied, one could use Bayesian inference to adaptively update the value of $L$ as an estimate of $a$ is attained, eventually arriving at the value $L_{\text{opt}}$. 

%\begin{figure*}[hbtp]
%\centering
%\includegraphics[width=\linewidth]{Figures/secondimage_v01.pdf}
%\caption{
%(a) The quantum advantage in estimation the absorbance as a function of the experimental transmission value $\gamma$. The value of $\gamma$ provides the value of the length \emph{independent} loss, applied twice for loss occurring before and after the sample of interest.
%(b) The Fisher information for a fixed $a=1$ as a function of length for different input probe Fano factor factors $\sigma_{{\psi}}$. We see that the maximal Fisher information decreases with increasing $\sigma_{{\psi}}$. The optimal length also increases with $\sigma_{{\psi}}$.
%(c) The Fisher information on the transmission $\epsilon$ where the experiment applied $i$ applications of the sample (such that total transmission is given by $\eta = \epsilon^i$). We see that this multi-pass strategy is a discrete case of the absorbance estimation following a similar trend to plot (b). We note that here we have plot a continuous line to demonstrate the similarities between (c) and (b), however $i$ should in principal only be able to take discrete integer values. 
%\label{fig:FImultiplots2}}
%\end{figure*} 

%The optimal length i
In the classical case, $L_{\text{opt}}$ is independent of $\gamma$ and so is independent of facet loss. When both schemes are 
%allowed to 
operated with
%at 
their respective 
%optimal length values
$L_{\text{opt}}$, the quantum advantage is found to be independent of 
%the co-propagating loss value 
$\beta$. The absolute Fisher information for each scheme is reduced as a co-propagating loss is introduced (non-zero $\beta$) but they are reduced by the same amount such that the ratio, $\mathcal{Q}(a)$, remains unchanged. This is not the case for the value of $\gamma$, which has a more detrimental effect on $\mathcal{F}_Q(a)$ than it does on $\mathcal{F}_C(a)$.
%to the Fisher information provided by the quantum scheme than in the classical case. 
Supplementary Material B~\cite{supplmaterial} displays how the quantum advantage is changed as $\gamma$
%facet transmission 
is varied. 

%\subsection{Arbitrary States}
%Thus far we have considered the Fisher information provided by coherent and Fock state inputs, as these provide the optimal strategies for classical and quantum cases. 
We now expand our
%the 
analysis to consider states more general than Fock and coherent states  $\ket{\psi}$, with
%of 
arbitrary Fano factor $\sigma_{{\psi}} = \Var{ N_{{0}}} / \bar{{N}}_{{0}}$ where $\Var{ N_{{0}}}$ and $\bar{{N}}_{{0}}$ are the input photon number variance and mean respectively. This allows the estimation capabilities of any light source with any statistics to be found. In Supplementary Material C~\cite{supplmaterial} we investigate estimation bounds in total \emph{absorption} estimation when considering general input states of light and find that the Fisher information on the transmission $\eta$ is given by
\begin{equation}
\mathcal{F}_{\psi}(\eta) = 1/{\left(\eta^2 \sigma_{{\psi}} + \eta(1-\eta)\right)},\label{FisherPSI}
\end{equation}
where for simplicity we have assume $\eta_l=1$. Following the same analysis as the Fock and coherent state inputs, where the Fisher information in $\eta$ is related to the Fisher information of $a$ using Eq.~\ref{eq:BLlaw_reparamFI}, the Fisher information is found to be
\begin{equation}
\mathcal{F}_{\psi}(a) = {L^2}/{\left(\sigma_{{\psi}} + \exp(a L) -1\right)}.
\end{equation}

%We can see that f
For the specific cases of the coherent and Fock states ($\sigma_{{\psi}} = 1$ and $0$ respectively), 
$\mathcal{F}_{\psi}(a)$ returns to $\mathcal{F}_{C}(a)$ and $\mathcal{F}_{Q}(a)$.
%the Fisher information value returns to the same as predicted in the early part of this paper. 
%The Fisher information for different $\sigma_{{\psi}}$ and fixed $a=1$ is shown in the \tr{supplementary material}. 
The dependence of the optimal length on $\sigma_{{\psi}}$ is found analytically to be 
\begin{equation}
L_{\text{opt}}^{{\psi}} = \left(\mathcal{W}\!\left[{2(\sigma_{{\psi}}-1)}/{e^2} \right] +2\right)/a.
\end{equation}
%The optimal length as a function of $\sigma_{{\psi}}$ for fixed $a=1$ is shown in \tr{the supplementary material Fig.~XXX}.

%\subsection{Multipass}

The general features of absorbance estimation also appear 
%when one investigates how 
in multipass or multi-application strategies 
%change the estimation capabilities 
for quantum and classical light. Specifically, the
%The
freedom of multiple passes acts like a discrete version of optimising the length and allows the classical scheme to reach close to the estimation capabilities of the quantum strategy. We show this
%For comparison, 
in Supplementary Material D~\cite{supplmaterial} by
%we 
revisiting work by Birchall \textit{et al.} who investigated multipass strategies in lossy phase estimation~\cite{birchall2017quantum} and loss estimation~\cite{birchall2019quantum}. We investigate the problem of estimating a sample transmission $\epsilon$ 
that is interrogated multiple times by
%but where the experimentalist is 
either 
%allowed to 
applying $\epsilon$
%this loss 
%an integer 
$i$ 
%multiple 
times to the beam (with $i$
%through having many 
copies of 
%the sample
$\epsilon$), or 
%is allowed to 
by passing light through the same sample $i$ times, such that the total loss on the optical beam is $\eta = \epsilon^i$. 

\begin{figure}[hbtp]
\centering
\includegraphics[width=\linewidth]{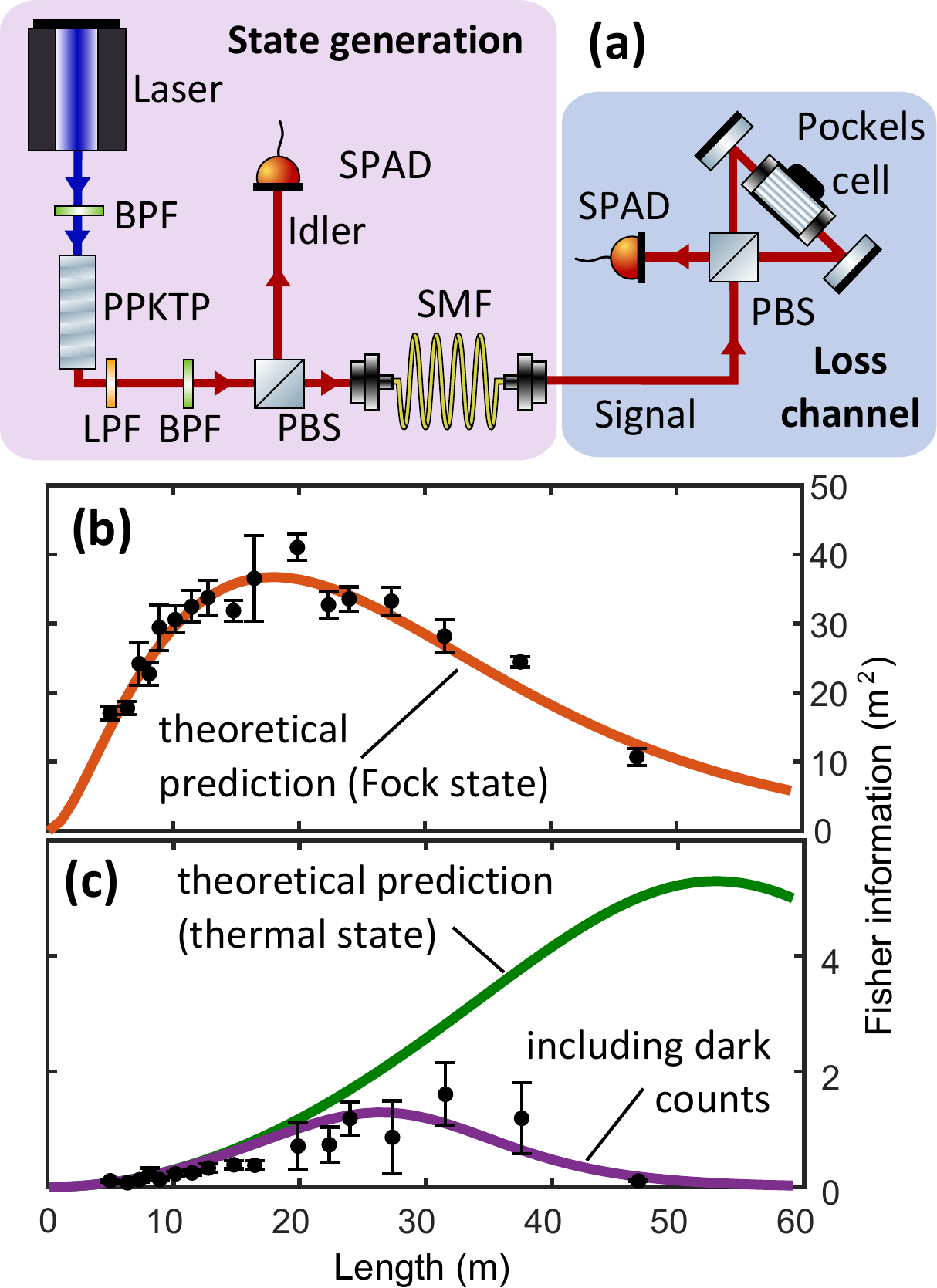}
\caption{\label{fig:experiment}
(a) Experimental setup to measure optimisation of sample length
(b) Results for pair-photon source
(c) Results for single-armed source 
%both 
with and without dark counts included in the theoretical analysis.} 
\end{figure}

%\section{Experiment}
We experimentally
%now present work 
demonstrate
%ing 
the length dependent optimisation that can be performed for both quantum (single photon) and noisy (thermal) sources of light using the experimental setup illustrated in
%. 
Fig.~\ref{fig:experiment}(a). 
%displays the setup of the experiment. 
A collinear type II spontaneous parametric down conversion (SPDC) source (periodically-poled potassium titanyl phosphate/PPKTP crystal) is pumped by a 3~mW 404~nm continuous-wave (CW) laser and spontaneously produces correlated photon pairs at 810~nm that have orthogonal polarisations. The spectral output of the crystal is controlled by varying the temperature using an oven. The pair of photons are spatially separated at a PBS into the `signal' and `idler' channels. The signal photon is detected by a single photon avalanche photodiode (SPAD) which heralds the presence of the other photon, producing a single photon Fock state. Long pass (LPF) and band pass filters (BPF) are used to filter out the pump. 

The sample of varying loss is implemented using a Pockels cell modulator composed of two lithium niobate crystals. When inactive, these crystals rotate the polarisation of the photon by 90 degrees but cause no rotation when activated. When used in conjunction with a polarising beamsplitter (PBS), this applies a variable loss to the incoming photon. At the wavelength of interest, the halfway voltage of the device is 200~V and so the crystal is driven by a high voltage driver. To make the loss independent to the photon input polarisation (which is arbitrary due to the single-mode optical fibre prior to the loss channel), 
we used the Sagnac configuration from~\cite{javier2017sub}.
%it is configured in a Sagnac configuration (\tr{see~\cite{sabines2018thesis} for details}). 
The transmission of the idler photon's path with the switch fully open is 38\%, 
%which 
including
%es 
detection efficiency, and thus we apply $\gamma = \sqrt{0.38} = 0.62$ and $\beta=0$ to the following analysis. The experiment produced approximately 14 k coincidences/s. 

Optimisation of 
%the length of an absorbing sample 
$L$
is observed 
%can be simulated 
by applying the inferred total loss from an object 
%fictitious sample 
with absorbance $a=1$ m$^{-1}$ with varying $L$.
%lengths. 
For example, for $a=1$ m$^{-1}$, 
%the lengths of 
$L=\{1,3,5\}$ m applies
%y the 
losses $\eta=\exp{(-a L)} = \{0.37,0.14,0.05\}$, that we implement with
%which can be implemented using a variable voltage applied to 
the Pockels cell. By setting each $\eta$
%the loss 
and then taking measurements to estimate 
%the value for 
$a$ (given that we have prior knowledge of $L$), 
%it is possible to gain 
we measure the
statistics of the noise on the estimate of absorbance and hence estimate
%the Fisher information via 
$\mathcal{F}(a) = 1/\Var{\hat{a}}$, where $\Var{\hat{a}}$ denotes the variance on the estimates of the parameter $a$. 

Estimates of $a$ are found by using the estimator 
\begin{equation}
\hat{a} = - \log \left( {N_{cc}}/{\gamma N_S} \right)/L, \label{eq:aestimator}
\end{equation}
where $N_{cc}$ is the number of coincidences between the signal and idler detectors, and $N_S$ is the total number in the signal channel. This is an adaptation of the estimator used 
%commonly in 
for absorption estimation with pair sources and photon counting~\cite{javier2017sub}. Eq.~\ref{eq:aestimator} is only valid for the Fisher information per incident photon metric where instrumental loss occurring before and after the sample affect $\mathcal{F}$ in the same fashion, and therefore do not need to be considered independently.
% to make it suited for absorbance estimation. 
 A total of 500 estimates for each setting
% value 
 of absorbance were found using a coincidence window of 0.5 seconds. These were separated into five groups of 100 estimates, and the variance of each group computed each providing a Fisher information estimate. The mean of these are shown in Fig.~\ref{fig:experiment}(b) and error bars are computed using the standard error of these Fisher information estimates. These 
% which
 show good agreement with theoretical predictions for the Fock state strategy.
% pair photon case (Fock state). 

%XXX -\tr{dealing with the intrinsic loss of the system}

%\subsection{Single Arm and Dark Counts}
By disregarding
%ignoring 
the idler photons 
%arm it is possible to 
we experimentally test the estimation capabilities of a noisy (thermal) light source as a single arm of a pair photon source is a thermal state~\cite{blauensteiner2009photon}. In this case, the estimator for the absorbance is changed to
\begin{equation}
\hat{a} = - \log \left[ \left({N_I - {N}_{DC}}\right)/{\bar{N}_0} \right] / L,
\end{equation}
where $N_I$ is the number of idler photons detected,
% counts in the idler arm, 
 $N_{DC}$ is the mean number of dark counts in the idler detector, and $\bar{N}_0$ is the mean number of input photons, 
% (which can be 
 found prior by applying no loss.
% ).

%R
The experimental results
%presented 
in Fig.~\ref{fig:experiment}(c)
%(b)~\&~(c) demonstrate some deviation of experimental results of the single arm data 
deviate from the theory given by Eq.~\ref{FisherPSI}. We attribute this to
%from what is expected from theoretical predictions . 
%It is hypothesised that this is the result of 
dark counts from the detectors, which are particularly detrimental for the thermal state (single arm) strategy as the counts from the lossy arm are used to estimate $\eta$.
%in the estimate of the loss (
This is in contrast to the more robust
%this is not true for the 
coincidence estimator (Eq.~\ref{eq:aestimator}) where only the singles in the signal channel are considered.
%). 
By adding dark counts into the analysis of the Fisher information (Supplementary Material E~\cite{supplmaterial}) we are able to reconcile the difference between the theoretical prediction and the experimental results seen in Fig.~\ref{fig:experiment}(c).
We estimate the
%The 
$\sigma_{\psi}$ value for the idler path
%arm can be estimated 
by 
%considering
measuring the photon number variance and mean when there is no sample present and correcting for the inherent loss of the channel (see Supplementary Material F~\cite{supplmaterial}). This gave a predicted $\sigma_{\psi} =826$. 
%Results of this single channel estimation can be seen in Fig.~\ref{fig:experiment}(b) and expanded in Fig.~\ref{fig:experiment}(c).

%We see that the single channel data does not show good agreement with theoretical predictions. Note that the single channel estimation of his highly susceptible to dark counts when compared with the coincidence schemes since this is the only estimator to explicitly consider the singles counts in the idler channel which can be low due to the sample loss (as opposed to the singles in the signal channel which are much higher and the coincidences across the channels that are time-correlated and so more tolerant to dark counts). 

Including dark counts provides an estimator information per incident photon of
\begin{equation}
I(\eta) =  \left[\Var{ N_{DC} }/{\bar{N}_0} + \eta^2 \sigma_{\psi} + (1-\eta) \eta\right]^{-1},
\end{equation}
where $\Var{ N_{DC} }$ is the variance of the detector dark counts. We expect this estimator information bound to be optimal (and hence equal to the Fisher information), but have no rigorous proof of this. $\Var{N_C}$ in this experiment was measured to be 518 $s^{-2}$. 

%\section{Conclusion}
%In this paper w
We have shown how experimental optimisation of the sample length $L$ can offer significant advantages in precision for both quantum and classical inputs in absorbance estimation. We find that for cases where this optimisation is possible, the quantum advantage is restricted to $1.2$ at most. We have derived optimal operating conditions for a number of experimental variations including additional experimental loss, input states with arbitrary photon number statistics, and detector dark counts. The experimental implementation presented demonstrates that $L$ can be optimised for Fock and thermal states, suggesting that for such an experiment the quantum advantage would be limited. These results not only have implications for future quantum sensors designed for measuring absorbance but can also be applied to optimise current classical sensors using laser or thermal light. 

The existence of experimental optimisation strategies for absorbance, phase~\cite{birchall2017quantum}, and loss estimation~\cite{birchall2019quantum,supplmaterial}, suggest that in general the existence of a free optimisation parameter can be a powerful tool for increasing the achievable measurement precision. Future analysis and experiments should consider this in order to correctly predict the advantage provided by quantum strategies. Efforts towards providing practical advantages using quantum states of light can now be focussed towards applications where such optimisation strategies are difficult to implement, such as imaging or very weakly absorbing samples.

\bibliographystyle{apsrev4-1} % Tell bibtex which bibliography style to use
%\bibliography{EAthesis_BLestimation}
%

\onecolumngrid
\newpage
\section{Supplementary Material}
\subsection{Supplementary Material A: Total Absorption ($\eta$) Estimation with Loss}
It has previously been shown~\cite{whittaker2017absorption,javier2017sub} that for transmission estimation, where the goal is to estimate a transmission $\eta$ defined by the input $I_0$ and output intensity $I = \eta I_0$, the Fisher information per incident photon for a coherent and Fock state input are given respectively to be
\begin{eqnarray}
\mathcal{F}_C(\eta) &=  \frac{1}{\eta}, \label{eq:FishClass}\\
\mathcal{F}_Q(\eta) &= \frac{1}{\eta (1- \eta)} \label{eq:FishQuant},
\end{eqnarray}
where $\mathcal{F}$ bounds the variance on any unbiased estimate of $\eta$ via 
\begin{equation}
\mathcal{F}(\eta) \leq \frac{1}{\Var{\hat{\eta}}},
\end{equation}
where $\hat{\eta}$ denotes an estimator of the parameter $\eta$. We note that these bounds were derived for the case where no experimental loss (other than for the sample $\eta$) exists. Here we derive how these bounds are changed by the introduction of experimental loss is also applied to the channel (e.g. by inefficient detectors). This loss is characterise by a second transmission parameter $\eta_l$ such that the input and output intensities are defined by $I = \eta \eta_l I_0$.

Assuming no change to the photon energy through the experiment, the output photon number from the lossy channel is given by 
\begin{equation}
\bar{N} = \eta \eta_l \bar{N}_0
\end{equation}
where $\bar{N}_0$ and $\bar{N}$ are the mean input and output photon numbers respectively. Considering the estimator $\hat{\eta}=N/(\eta_l \bar{N}_0)$ (which has previously been show to saturate the Fisher information bound~\cite{whittaker2017absorption,javier2017sub}) and using the error propagation formula: $\Var{\hat{\eta}} = \left(\frac{\partial \eta}{\partial N}  \right)^2 \Var{N}$ we can relate the variance on any estimate of $\hat{\eta}$ to the output photon number variance and find
\begin{equation}
\Var{\hat{\eta}} = \frac{\Var{N}}{(\eta_l \bar{N}_0)^2}.\label{eq:VarEtaToVarN}
\end{equation} 
For an input coherent state $\ket{\alpha}$ ($|\alpha|^2 = N_0$) input into a lossy channel, the output state is given by $\ket{\sqrt{\eta_l \eta} \alpha}$ and therefore the photon number statistics at the output follow a Poisson distribution with a characteristic variance equivalent to the it's mean $\Var{N}_C = \eta_l \eta N_0$. Similarly, a Fock state input will have output statistics defined by a Binomial distribution~\cite{whittaker2017absorption} with number of trials $N_0$ and success probability (chance of being transmitted) $\eta \eta_l$. As a result, for a Fock state --- the optimal quantum strategy --- the output photon number variance is given by $\Var{N}_Q = \eta_l \eta (1 - \eta_l \eta) \bar{N}_0$. 

Combining the classical and quantum output photon number variances with Equation~\ref{eq:VarEtaToVarN} and using $\mathcal{F}'(\eta) = 1/(\Var{\hat{\eta}} \bar{N}_0)$ we find
\begin{eqnarray}
\mathcal{F}_C(\eta) &=  \frac{\eta_l}{\eta},\\
\mathcal{F}_Q(\eta) &= \frac{\eta_l}{\eta (1- \eta \eta_l)},
\end{eqnarray}
as defined in the main text. 

\subsection{Supplementary Material B: Quantum Advantage as a Function of the Loss Parameter $\gamma$}
\begin{figure}[hbtp]
\centering
\includegraphics[width=0.6\textwidth]{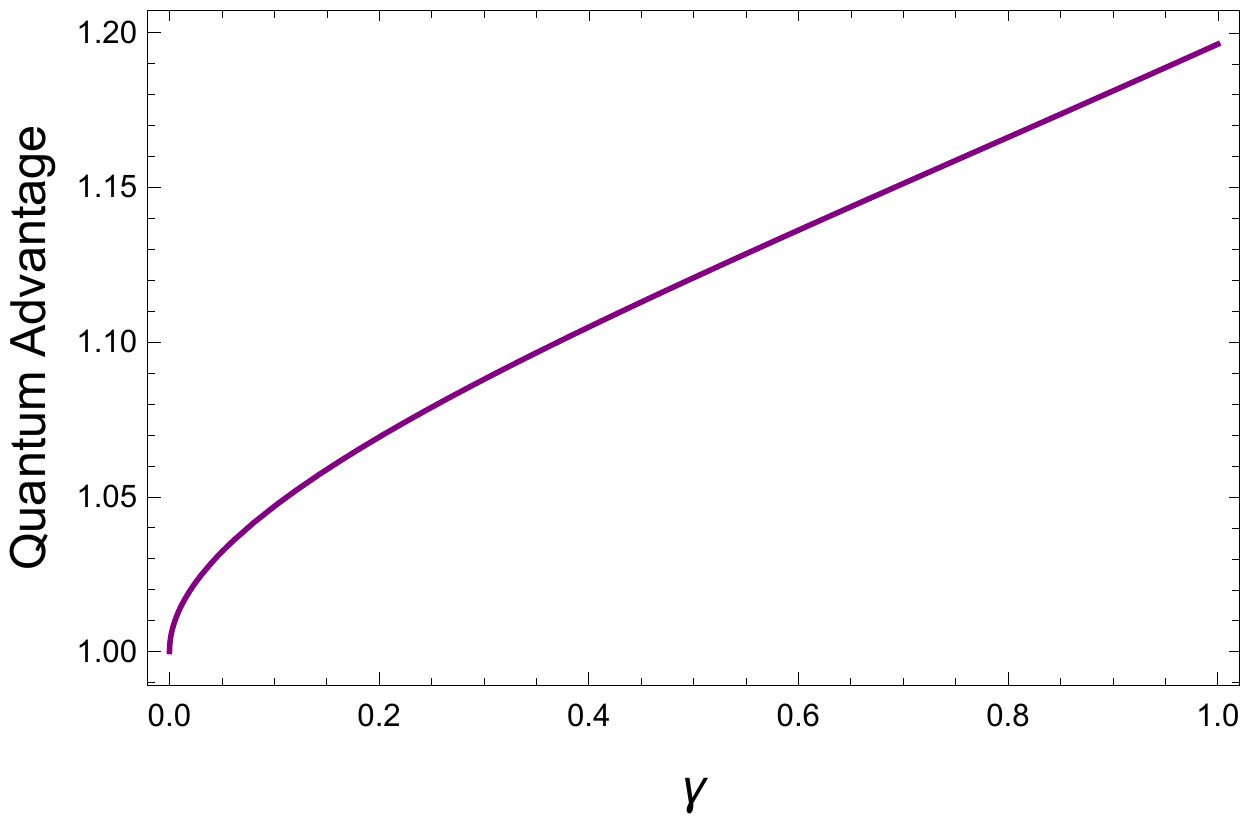}
\caption{The quantum advantage in estimation the absorbance as a function of the experimental transmission value $\gamma$. The value of $\gamma$ provides the value of the length \emph{independent} loss, applied twice for loss occurring before and after the sample of interest.\label{fig:QAWithgamma}}
\end{figure}

Figure~\ref{fig:QAWithgamma} demonstrates how the quantum advantage varies as a function of the length independent facet transmission $\gamma$ (defined in the main text). We see that unlike the length dependent loss factor $\beta$ the quantum advantage offered by Fock states is reduced as the transmission $\gamma$ is reduced. 

\subsection{Supplementary Material C: Absorption Estimation with Arbitrary Input States}

%- Including plot showing The Fisher information for different XX and fixed $a = 1$ is shown in the supplementary material.
%- Including plot showing The optimal length as a function of xx for fixed $a = 1$ is shown in the supplementary material Figure XXX.

Here we derive the Fisher information attained on $\eta$ when arbitrary states, $\ket{\psi}$ characterised by the Fano factor $\sigma_{{\psi}} = \Var{ N_{{0}}} / \bar{N}_0$ where $\Var{ N_{{0}}}$ and $\bar{N}_0$ are the input photon number variance and mean respectively, are incident on the absorbing medium.

Let $X({N}_0)$ define the (arbitrary) photon number distribution describing the input state of light used to probe a sample of transmission $\eta$ where ${N}_0$ is the input photon number. As we are interested in linear absorption processes, where the loss acts independently on each incoming photon, passing this light through the sample applies a Binomial distribution to the input distribution and results in the output compound distribution %$\mathcal{X}_B(N|\eta)$ where
\begin{equation}
 \mathcal{X}_B(N|\eta) = \sum_{{N}_0=0}^{\infty}X({N}_0)B(N|{N}_0,\eta) , 
\end{equation}
where $B(N|{N}_0,\eta)$ defines a binomial distribution providing the probability of measuring $N$ sucesses with ${N}_0$ trials and a chance of success of $\eta$. The expectation values and variance of the Binomial distribution 
\begin{equation}
B(N|{N}_0,\eta) = {{N}_0 \choose N} \eta^N (1-\eta)^{{N}_0-N},
\end{equation}
can be calculated to be:
\begin{align}
\mathds{E}_B[N] &\defeq \sum_{N=0}^{\infty}N B(N|{N}_0,\eta) = {N}_0\eta, \\
\mathds{E}_B[N^2] &\defeq \sum_{N=0}^{\infty}N^2 B(N|{N}_0,\eta) \nonumber \\ 
&= {N}_0 \eta +{N}_0^2 \eta^2-{N}_0 \eta^2 \\
&= \eta^2 {N}_0^2  + \eta(1-\eta){N}_0, \\
\Varr_B(N) &= \mathds{E}_B[N^2] - \mathds{E}_B[N]^2\\
&= {N}_0 \eta -{N}_0 \eta^2 = {N}_0 \eta(1-\eta),
\end{align} 
where $\mathds{E}_F[x]$ denotes the expectation value of $x$ for the distribution $F$. For the compound distribution $\mathcal{X}_B(N|\eta)$:
\begin{align}
\mathds{E}_{\mathcal{X}_B}[N] &\defeq \sum_{N=0}^{\infty}N \mathcal{X}_B(N|\eta) \nonumber \\
&= \sum_{N=0}^{\infty}N \sum_{{N}_0=0}^{\infty}X({N}_0)B(N|{N}_0,\eta) \nonumber \\
&= \sum_{{N}_0=0}^{\infty}X({N}_0) \mathds{E}_B[N] \nonumber \\
&= \sum_{{N}_0=0}^{\infty}X({N}_0) {N}_0\eta = \eta \mathds{E}_X[{N}_0], \nonumber\\
\mathds{E}_{\mathcal{X}_B}[N^2] &\defeq \sum_{N=0}^{\infty}N^2 \sum_{{N}_0=0}^{\infty}X({N}_0)B(N|{N}_0,\eta) \nonumber \\
&= \sum_{{N}_0=0}^{\infty}X({N}_0) ({N}_0^2 \eta^2+{N}_0( \eta - \eta^2)) \nonumber \\
&= \eta^2\mathds{E}_X[{N}_0^2]+\eta(1-\eta)\mathds{E}_X[{N}_0], \nonumber\\
\Varr_{\mathcal{X}_B}(N) &= \eta^2\mathds{E}_X[{N}_0^2]+\eta(1-\eta)\mathds{E}_X[{N}_0] - \eta^2\mathds{E}_X[{N}_0]^2 \nonumber\\
&= \eta^2 \Varr_X({N}_0) + \eta(1-\eta)\mathds{E}_X[{N}_0]. \label{eq:XBnumberVariance}
\end{align}
\begin{figure*}[hbtp]
\centering
\includegraphics[width=0.9\textwidth]{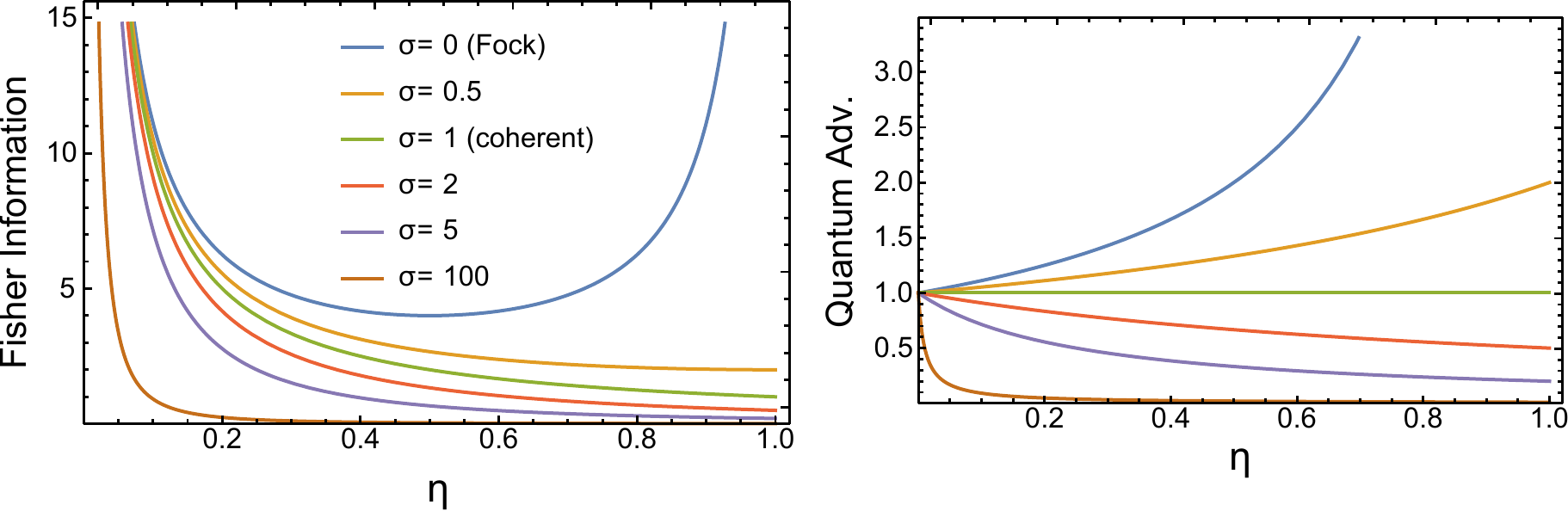}
\caption{The Fisher information on the parameter $\eta$ per incident photon $\mathcal{F}(\eta)$ (left) and quantum advantage (right) for states with varying Fano factors ($\sigma = \frac{\Var{{N}_0}}{\bar{{N}}_{0}}$). We observe that reducing the $\sigma$ provides an increase in the Fisher information available with that input state. \label{fig:qsensing_arbStateFIandQA}}
\end{figure*}
From Equation~\ref{eq:XBnumberVariance} one can deduce that for an input photon number distribution of $X({N}_0)$, the photon number variance of the light after the sample (where the sample transmission coefficient is $\eta$) is given by
\begin{equation}
\Varr_{\mathcal{X}_B}(N) = \eta^2  \Var{{N}_0} + (1-\eta)\eta \bar{{N}}_{0}, \label{eq:ARBnumberVariance}
\end{equation}
where $\mathds{E}_X[N_0]=\bar{{N}}_{0}$ is the mean number of input photons and $\Var{{N}_0}$ is the variance. Using the estimator $\hat{\eta} = N/\bar{N}_0$ allows the variance on the estimator to be computed using the relationship for the propagation of errors: $\Var{\hat{\eta}} = \left(\frac{\partial \hat{\eta}}{\partial N}  \right)^2 \Varr_{\mathcal{X}_B}(N)$:
\begin{align}
\Var{\hat{\eta}} &= \frac{\Var{N}_{\mathcal{X}_B}}{\bar{{N}}_{0}^2} \nonumber \\
&=  \frac{\eta^2 \Var{{N}_0} + (1-\eta)\eta \bar{{N}}_{0}}{\bar{{N}}_{0}^2} \nonumber \\
&= \frac{1}{\bar{{N}}_{0}} \left( \eta^2 \left( \frac{\Var{{N}_0}}{\bar{{N}}_{0}} \right) + (1-\eta)\eta \right)\nonumber \\
&= \frac{1}{\bar{{N}}_{0}} \left( \eta^2 \sigma  + (1-\eta)\eta \right).\label{eq:Xoptimality}
\end{align}
From this we find that the Fisher information on the parameter $\eta$ per incident photon is given by
\begin{equation}
\mathcal{F}(\eta) = \frac{1}{ \eta^2 \sigma  + (1-\eta)\eta}.
\end{equation}

Note, that for the coherent state case ($\sigma = 1$), we arrive at the same variance as discussed previously. Also, the Fisher information is  maximised for the case where $\sigma = 0$, which is true for Fock states. Therefore, we confirm Fock state is an optimal probe state for a linear loss channel, as found previously~\cite{adesso2009optimal,fujiwara2004estimation}. Figure~\ref{fig:qsensing_arbStateFIandQA} displays the relationship with the input photon number $\sigma$ value and the Fisher information.

\subsection{Supplementary Material D: Multipass Strategies}
%The general features of absorbance estimation also appear when one investigates how multipass or multi-application strategies change the estimation capabilities for quantum and classical light. For completeness, here we look at the problem of estimating a sample transmission $\eta$ but where the experimentalist is either allowed to apply this loss multiple times to the beam (through having many copies of the sample), or is allowed to pass the light through the same sample multiple times (see Figure~\ref{fig:qsensing_multipasssketch}). This type of estimation procedure has been previously studied by Birchall \textit{et al.} who investigated multipass strategies in lossy phase estimation~\cite{birchall2017quantum} and loss estimation~\cite{birchall2018thesis}.

\begin{figure}[hbtp]
\centering
\includegraphics[width=0.9\textwidth]{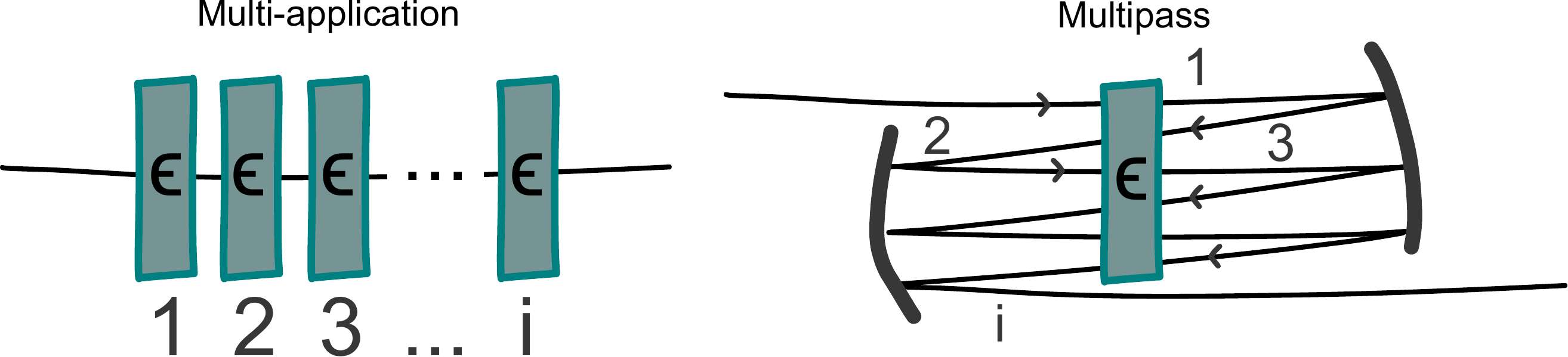}
\caption[Sketch of practical ways to implement a multi-application or multipass strategy.]{Sketch of practical ways to implement a multi-application (left) or multipass (right) strategy. \label{fig:qsensing_multipasssketch}}
\end{figure}

The general features of absorbance estimation also appear when one investigates how multipass or multi-application strategies change the estimation capabilities for quantum and classical light. For completeness, here we look at the problem of estimating a sample transmission $\epsilon$ but where the experimentalist is either allowed to apply this loss multiple times to the beam (through having many copies of the sample), or is allowed to pass the light through the same sample multiple times (see Figure~\ref{fig:qsensing_multipasssketch}). This type of estimation procedure has been previously studied by Birchall \textit{et al.} who investigated multipass strategies in lossy phase estimation~\cite{birchall2017quantum} and loss estimation~\cite{birchall2019quantum}. 

\begin{figure}[hbtp]
\centering
\includegraphics[width = 0.5\textwidth]{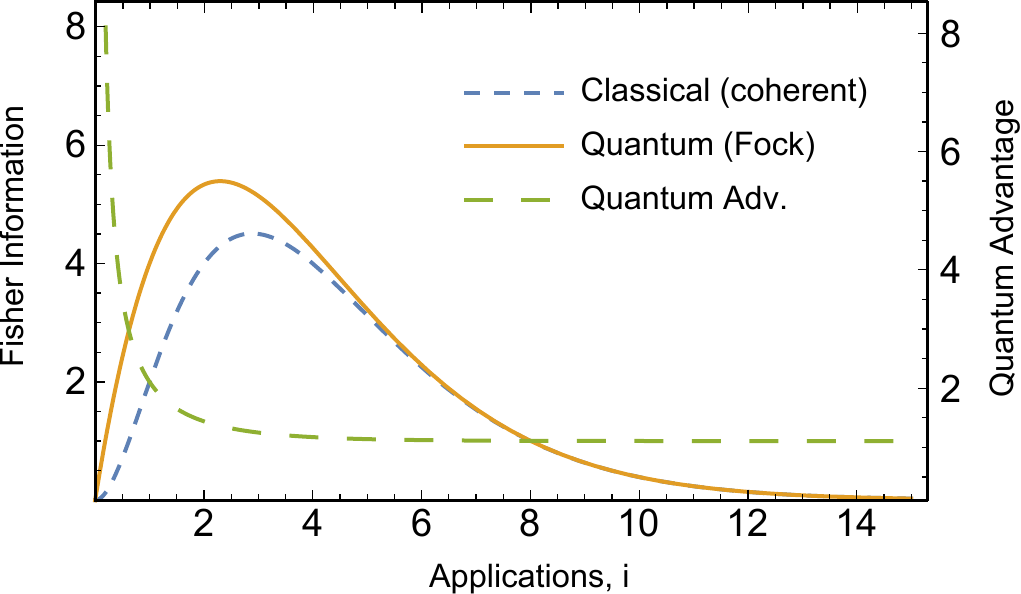}
\caption[Fisher information per incident photon $\mathcal{F}'(\epsilon)$ as a function of the number of passes $i$ for $\epsilon = 0.5$.]{Fisher information per incident photon $\mathcal{F}'(\epsilon)$ as a function of the number of passes $i$ for $\epsilon = 0.5$. Distribution shows that applying the loss three (two) times in the classical (quantum) case would provide the most precise estimate of $\epsilon$. We note that practically $i$ is limited to discrete values but we have plotted a continuous line to display the simarlarities of this result with that in the main text. \label{fig:qsensing_multipassFI}}
\end{figure}

Figure~\ref{fig:qsensing_multipasssketch} demonstrates the two implementations of a multipass or multiapplication strategy. In this case the total transmission of the optical beam is given by
\begin{equation}
\eta = \epsilon^i, 
\end{equation}
where $i$ is the number of applications of the sample $\epsilon$ or beam passes, and $\epsilon$ is the single-pass sample transmission. Using the same Fisher information propagation analysis used in the main text, where $\mathcal{F}(\epsilon) = \left( \frac{\partial \eta}{\partial \epsilon} \right)^2 \mathcal{F}(\eta)$, we find that for a coherent state the Fisher information per incident photon is found to be
\begin{equation}
\mathcal{F}(\epsilon)_{\ket{\alpha}} = \frac{i^2}{\epsilon^{2-i}},
\end{equation}
and for the Fock state
\begin{equation}
\mathcal{F}(\epsilon)_{\ket{N_0}} = \frac{i^2}{\epsilon^{2-i}(1-\epsilon^i)}.
\end{equation}
The value of the Fisher information for fixed $\epsilon = 0.5$ as a function of applications $i$ is shown in Figure~\ref{fig:qsensing_multipassFI}.

Figure~\ref{fig:qsensing_multipassFI} displays very similar behaviour as the Beer-Lambert law absorbance case that was discussed in the main text. In this case, applying the loss sample more that once can increase the precision on the estimate of it's value. This is similar to increasing the length of the absorbing sample for the absorbance case.

\begin{figure}[hbtp]
\centering
\includegraphics[scale=0.85]{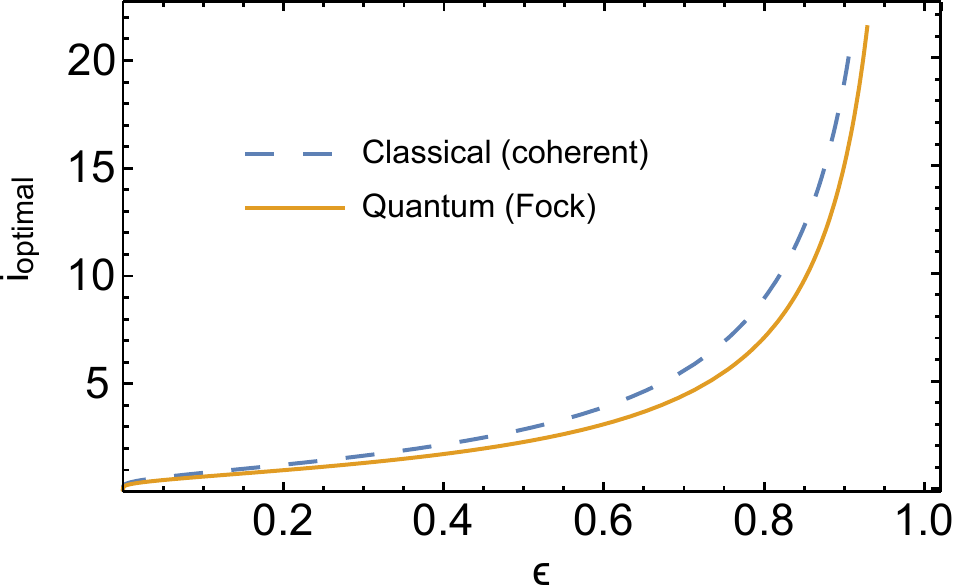}
\caption[Optimal values of $i$ (number of passes or implementations of the loss $\epsilon$) for maximising $FI'(\epsilon)$ for classical and quantum cases.]{Optimal values of $i$ (number of passes or implementations of the loss $\epsilon$) for maximising $FI'(\epsilon)$ for classical and quantum cases. \label{fig:qsensing_multipassOptimali}}
\end{figure}

The optimal value of $i$ can be found by finding the location of the maxima of the Fisher information functions in Figure~\ref{fig:qsensing_multipassFI} using the same differential approach as the main paper. This produces the value $i^C_{optimal} = - \frac{2}{\log{\epsilon}}$ and $i^Q_{optimal} = - \frac{W(\frac{2}{e^2})-2}{\log{\epsilon}}$ for the classical and quantum cases respectively. The optimal values for the number of passes are shown in Figure~\ref{fig:qsensing_multipassOptimali}. This fixes the optimal total absorption $\eta$ to be around 13.5\% and 20.3\% in the classical and quantum cases respectively (for all values of $\epsilon$). 

We see from Figure~\ref{fig:qsensing_multipassFI} that the multipass strategy holds many similarities with the Beer-Lambert law case in the main text, where increasing the interaction of the light with the sample can increase the information provided on the loss, even when this results in more total loss being applied to the beam. In this sense, the multipass strategy can be thought of as a discrete version of the Beer-Lambert law (since the number of applications of the sample are limited to integer values).

\subsection{Supplementary Material E: Adding Dark Counts to Information Analysis}
Results presented in the main text demonstrate deviation of experimental results of the single arm data from what is expected from theoretical predictions. It is hypothesised that this is the result of dark counts from the detectors, which are particularly detrimental for the single arm strategy as the counts from the lossy arm are used in the estimate of the loss. This is not true for the coincidence estimator $\hat{\eta} = N_{signal}/N_{CC}$ where only the singles in the signal channel are considered. Here we look in more detail at the effect of dark counts on the estimation capabilities of a single-arm measurement in a multipass scheme in an effort to reconcile the difference between the theoretical prediction and experimental results. 

For a single-arm measurement, estimates of $\eta$ are calculated using the estimator
\begin{equation}
\hat{\eta} = \frac{N_C - \bar{N}_{DC}}{\bar{N}_0} \label{eq:qsensing_darkcounts1}
\end{equation}
where $N_C$ is number of detector counts in a sample window, $\bar{N}_{DC}$ is the average number of dark counts over many sample windows, and $\bar{N}_0$ is the average number of input photons. The number of counts on the detector is the sum of dark counts in the sample window and the number of incident photons, $N_i$, $N_C = N_{DC} + N_{i}$. Because of this relationship, and that $N_{i}$ and $N_{DC}$ are independent variables, the variance on $N_C$ is just the sum of the individual variances of $N_{i}$ and $N_{DC}$
\begin{equation}
\Var{ N_C } =  \Var{N_{i} } + \Var{N_{DC}}. 
\end{equation}
Using the standard error propagation formula
\begin{equation}
\Var{\hat{\eta}} = \left( \frac{\partial \hat{\eta}}{\partial N_C} \right)^2 \Var{N_C}
\end{equation} 
with Equation~\ref{eq:qsensing_darkcounts1}, we find
\begin{equation}
\Var{ \hat{\eta}} = \frac{\Var{ N_C} }{\bar{N}_0^2} = \frac{ \Var{ N_{i} } +\Var{N_{DC}}}{\bar{N}_0^2}.
\end{equation}
We can relate the detected incident photon number variance with the input (pre-sample) variance using Equation~\ref{eq:ARBnumberVariance}
\begin{equation}
\Var{ N_i } = \eta^2 \Var{N_0} + (1-\eta)\eta \bar{N}_0,
\end{equation}
which can then be used to give the variance on $\hat{\eta}$ for an arbitrary input states with $\sigma = \Var{N_0} / \bar{N}_0$
\begin{equation}
\Var{\hat{\eta}} = \frac{1}{\bar{N}_0} \left( \frac{\Var{ N_{DC} }}{\bar{N}_0} + \eta^2 \sigma + (1-\eta) \eta \right).
\end{equation}
This provides an estimator information per incident photon of
\begin{equation}
I(\eta) = \frac{1}{\bar{N}_0 \Var{\hat{\eta}}}   = \frac{1}{\frac{\Var{ N_{DC} }}{\bar{N}_0} + \eta^2 \sigma + (1-\eta) \eta}, 
\end{equation}
which simplifies to the Fisher information bounds for previous scenarios discussed in the main text. We expect this information bound to be optimal (and hence equal to the Fisher information), but have no rigorous proof of this.

\subsection{Supplementary Material F: Computing the $\sigma$ Value of a Source}
Here we show how to calculate the Fano factor $\sigma$ of a light source from measurement of the noise properties of the light after a channel of transmission $\eta_l$. This is useful when trying to estimate the properties of the source when taking measurements with inherent experimental loss. This technique is used in the experimental results of the main paper.

Equation~\ref{eq:ARBnumberVariance} relates the photon number variance after a loss channel to properties of the light before the channel:
\begin{equation}
\Var{N} = \eta_l^2 \Var{N_0} + (1-\eta_l)\eta_l \bar{N}_0.
\end{equation}
By dividing by the mean photon number after the channel, $\bar{N}$, we can relate the input sigma value $\sigma_0$ to the output one
\begin{align}
\sigma &= \frac{\Var{N}}{\bar{N}} = \frac{\Var{N}}{\eta_l \bar{N}_0}, \\
&= \frac{\eta_l^2 \Var{N_0} }{\eta_l \bar{N}_0} + \frac{(1-\eta_l)\eta_l \bar{N}_0}{\eta_l \bar{N}_0}, \\
&= \eta_l \sigma_0 + (1-\eta_l),
\end{align}
and therefore
\begin{equation}
\sigma_0 = \frac{\sigma + \eta_l - 1}{\eta_l}.
\end{equation}

Experimentally we measured a Fano factor $\sigma = 314.5$ where the channel transmission with no sample present was $\eta_l = 0.38$. This produces a value of $\sigma_0 = 826$ which is used as the correction factor in the main text.

%\section{Acknowledgements}
%We thank Sam Pallister and Patrick Birchall for their helpful discussions. This  work  was  supported  by  EPSRC  UK  Quantum  Technology  Hub  in Quantum Enhanced Imaging (EP/M01326X/1) and the Centre for Nanoscience and Quantum Information (NSQI). EJA acknowledges support form EPSRC Centre for Doctoral Training in Quantum Engineering (EP/L015730/1) and EPSRC Doctoral Prize Fellowship (EP/R513179/1). JCFM   acknowledges support  from  an  EPSRC  Quantum  Technology  Fellowship (EP/M024385/1) and an ERC starting grant ERC-2018-STG803665.

\end{document}